\documentclass[%
 aip,
 amsmath,amssymb,
 reprint,%
]{revtex4-1}

\usepackage{graphicx}
\usepackage{dcolumn}
\usepackage{bm}

\usepackage[utf8]{inputenc}
\usepackage[T1]{fontenc}
\usepackage{mathptmx}
\usepackage{etoolbox}

\makeatletter
\def\@email#1#2{%
 \endgroup
 \patchcmd{\titleblock@produce}
  {\frontmatter@RRAPformat}
  {\frontmatter@RRAPformat{\produce@RRAP{*#1\href{mailto:#2}{#2}}}\frontmatter@RRAPformat}
  {}{}
}%
\makeatother
\begin{document}

\preprint{AIP/123-QED}

\title[Exact solution for a general FJC polyelectrolyte model with up to Next Nearest Neighbour Interactions]{Exact solution for a general FJC polyelectrolyte model with up to Next Nearest Neighbour Interactions}
\author{Javier Orradre}
\email{j.orradre@ub.edu, fmas@ub.edu}
\affiliation{Department of Material Science and Physical Chemistry \& Institute of Theoretical and Computational Chemistry (IQTC), University of Barcelona, Barcelona, Catalonia, Spain}

\author{Pablo M. Blanco}%
\affiliation{Department of Material Science and Physical Chemistry \& Institute of Theoretical and Computational Chemistry (IQTC), University of Barcelona, Barcelona, Catalonia, Spain}
\affiliation{Department of Physics, Faculty of Natural Sciences, Norwegian University of Science and Technology (NTNU), Trondheim, Norway}

\author{Sergio Madurga}
\affiliation{Department of Material Science and Physical Chemistry \& Institute of Theoretical and Computational Chemistry (IQTC), University of Barcelona, Barcelona, Catalonia, Spain}

\author{Francesc Mas}
\affiliation{Department of Material Science and Physical Chemistry \& Institute of Theoretical and Computational Chemistry (IQTC), University of Barcelona, Barcelona, Catalonia, Spain}

\author{Josep Lluís Garcés}
\affiliation{Department of Chemistry, Physics and Environmental and Soil Sciences \& Agrotecnio, University of Lleida, Lleida, Catalonia, Spain}

\date{\today}

\begin{abstract}
This work presents a polyelectrolyte (PE) model based on a freely jointed chain (FJC) in which the beads are ionizable and the bonds are rigid. Electrostatic interactions considered are of the short range type up to next-nearest neighbours using a Debye-Hückel (DH) potential. Exact expressions for the conformational and protonation properties are obtained as a function of the $\mathrm{pH}$ using the transfer matrix method and the integration of the angular variables. For the finite cases with low $N$ values, relevant differences arise on both the protonation curve and the end to end distance due to the end effects. However, when increasing $N$ a rather rapid convergence is observed to the infinite chain limiting case.
\end{abstract}

\maketitle
\section{\label{Introduction}Introduction}

Linear weak polyelectrolytes (PE) are polymers whose chains contain ionizable chemical groups that can be electrically charged when dissolved in polar solutions. These chemical groups establish a protonation equilibrium with the environment and therefore, the charge of the PE is not a constant quantity. Fluctuations are due to conformational changes of the PE and other surrounding conditions such as pH, ionic strength, external forces, etc., and this phenomenon is commonly known as charge regulation\cite{Blanco2019b,Garces2017}. This singularity is the reason why weak PE play a crucial role in a variety of medical and biological applications including drug delivery\cite{Hartig2007}, colloidal stability\cite{Trefalt2016}, supramolecular chemistry\cite{Li2016} as well as many others.\cite{Blanco2023} Therefore, an accurate characterization of their properties such as the degree of ionization and the end to end distance is essential to understand their behaviour.

Here we present a simple FJC-based polyelectrolyte model with short range electrostatic interactions. Similar models have been previously used in other studies to describe the conformation and properties of PEs using Monte Carlo simulations and theoretical approximations.\cite{Marcus1954,Reed1992,Smits1993} The novelty in this case is the fact that the model presents the possibility to be exactly solved not only for the ionization degree but for the end to end distance too, which is a rather lacking result in most of the works.

In order to organize the deduction of the model as well as the results obtained, we have structured the article as follows. In section (\ref{ModelDescription}) the characteristics of the model are introduced as well as its corresponding free energy. Later, in Section (\ref{Method}) the transfer matrix methodology\cite{Flory1969,Chandler1987} is explained along with the modifications considered to include the conformational dependence in the energy treatment. Finally, in sections (\ref{FiniteChain}) and (\ref{InfiniteChain}), the degree of protonation and the end to end distance are found for the cases of the finite and infinite chains respectively.

\section{\label{ModelDescription}Model description}

\begin{figure}[h!]
\includegraphics[scale=0.4,trim= 0in 0in 0in 0in]{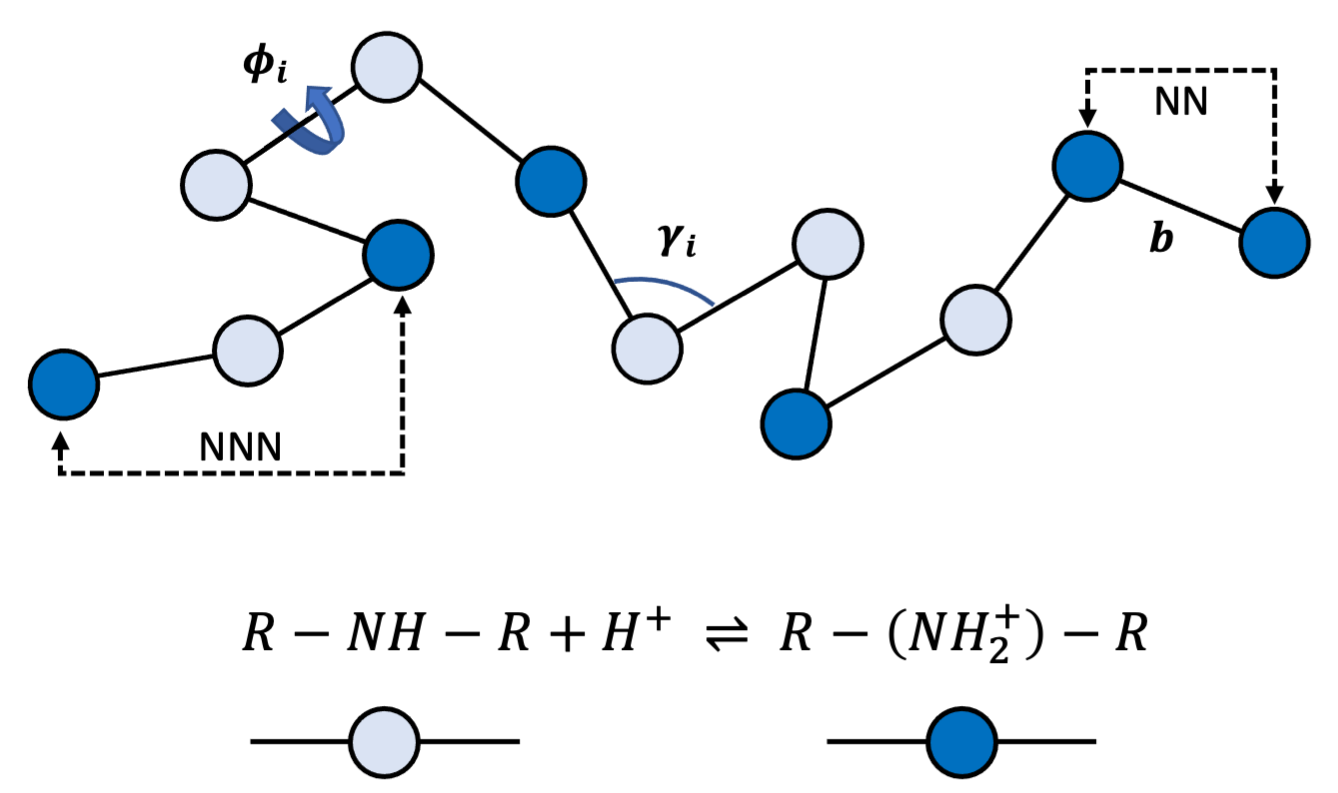}
\caption{PE model of an $N$ sites ionizable FJC connected with bonds of fixed length $b$. Each ionizable site undergoes a protonation equilibrium as described by the reaction above which is mediated by the reduced chemical potential $\mu$. The electrostatic interactions considered are of the DH type and take place between nearest (NN) and next-nearest (NNN) neighbouring sites, the latter depending on each bond angle $\gamma_{i}$. The dihedral angles $\phi_{i}$ can rotate freely.}
\label{Fig_Model}
\end{figure}

In this article our PE is represented as a chain of $N$ ionizable sites connected by $N-1$ bonds (Fig.\,\ref{Fig_Model}). Remember that each site is a coarse-grained depiction of the chemical ionizable group repeated along the chain. When the PE is uncharged, it behaves as a freely jointed chain (FJC) and the angles between consecutive bonds $\gamma_{i}\left(i=3,...,N\right)$ are statistically independent. The sites are identical and the protonation equilibria is governed by the reduced chemical potential $\mu_{i}\left(c\right)=\mu=\ln10\,(\mathrm{pH}-\mathrm{p}K)$, which has been considered independent of the conformational $c$ state. Moreover, two protonated sites only interact when they are nearest neighbours or next-nearest neighbours along the chain.\cite{Marcus1954,Smits1993} 

The distance $b$ of the bond is fixed and the nearest neighbour interaction energy is not conformational dependent $\epsilon_{i,i+1}\left(c\right)=\epsilon_{\mathrm{n}}$. One can choose the electrostatic interaction potential that considers most appropriate for the next-nearest neighbours as long as the form is of the type $\varepsilon_{i,i+2}=\varepsilon(r(\gamma_{\,i+2}))$. As an example, in this theoretical development all next and next-nearest neighbouring sites 
 will interact by means of the Debye-Hückel (DH) potential 
\begin{gather}
\nonumber\epsilon_{\mathrm{i,j}}\left(r\right)=\frac{1}{4\pi\varepsilon r}\exp\left(-r/\ell_{\mathrm{D}}\right)\\ with\,\,r=\begin{cases}
b & if\;j=i+1\\
r\left(\gamma\right)=b\sqrt{2\left(1-\cos\left(\gamma\right)\right)} & if\;j=i+2
\end{cases}\label{eq:epsNNN}
\end{gather}
where $\gamma$ is the angle between two consecutive bonds, $\varepsilon$ is the permittivity of the medium (which is assumed to be the same of water) and $\ell_{\mathrm{D}}(\textrm{nm})=0.304/\sqrt{I(\textrm{M})}$ is the Debye length ($\mathrm{H_2O}$ at 25\textdegree C), with $I$ being the ionic strength.

Thus, with all these considerations the free energy of the system reads
\begin{equation}
F\left(s,c\right)=\mu\sum_{i}s_{i}+\epsilon_{\mathrm{n}}\sum_{i}s_{i}s_{i+1}+\sum_{i}\epsilon_{\mathrm{nn}}\left(\gamma_{i+2}\right)s_{i}s_{i+2}
\label{eq:free energy contracted NNNN model}
\end{equation}
where the energy $\epsilon_{i,i+2}$ has been expressed as $\epsilon_{\mathrm{nn}}\left(\gamma_{i+2}\right)$ to emphasize the angular dependence, $s$ represents the protonation state of the PE ($s=\left\{s_{i}\right\}$) and $c$ its conformational state, which in this particular case is exactly determined as $c=\left\{\gamma_{i}\right\}$.

\section{\label{Method} Methodology}

\subsection{Standard treatment for constant energies}

To get to the solution for this model let us firstly consider a chain with $N$ sites and nearest neighbour interactions that are equal for all sites and do not depend on the configurations of the PE. Let us denote an empty and a protonated site with the symbols '$\circ$' and '$\bullet$' respectively. Here we will make extensive use of the transfer matrix approach, a rather standard method in statistical mechanics\cite{Chandler1987} and widely used in physical chemistry of polymers.\cite{Flory1969} The well known matrix for this system is given by\cite{Reed1992,Koper1996,Borkovec1996,Borkovec2006}
\begin{equation}
\mathbf{\mathbf{T_{\mathrm{n}}}}=\left(\begin{array}{cc}
1 & z\\
1 & zu_{\mathrm{n}}
\end{array}\right)\label{eq:TF nearest neighbor}
\end{equation}
where $\mu=-\mathrm{k}_{\mathrm{B}}T\ln z$ and $\epsilon_{\mathrm{n}}=-\mathrm{k}_{\mathrm{B}}T\ln u_{\mathrm{n}}$. The corresponding partition
function at constant pH (constant $\mu$) reads
\begin{equation}
\Xi=\mathbf{v}_{\mathrm{i}}\mathbf{T_{\mathrm{n}}}^{N}\mathbf{v_{\mathrm{f}}^{T}}\label{eq:PF nearest neighbor}
\end{equation}
with $\mathbf{v}_{\mathrm{i}}=\left(1,0\right),\mathbf{v}_{\mathrm{f}}=\left(1,1\right)$. The first column in $\mathbf{T}$ represents the addition of an empty site ($\circ$) and the Boltzmann factors are 1 since there is no increase in the total free energy, neither by protonation nor by
interactions. However, the second column represents the addition of a protonated site ($\bullet$) provided that the previous site is deprotonated (first row, increase in free energy due to protonation of the current site, $z$) or the previous site is protonated (second
row, increase in free energy due to the protonation of the current site and the nearest neighbour interaction with the preceding protonated site, $zu_{\mathrm{n}}$). 

In order to account for the next nearest-neighbour interactions, we have to add pairs of sites (instead of only one site, to consider the history of the $i-2$ site) such as the first site of the new pair must coincide with the second site of the previous pair. The pairs can adopt four possible states ($\circ-\circ$, $\bullet-\circ$, $\circ-\bullet$, $\bullet-\bullet$) and a 4x4 transfer matrix is needed. Moreover, some combinations are impossible and a zero Boltzmann factor must be assigned to them. For instance, one can not add the pair $\bullet-\circ$ (first site $\bullet$) to the pair $\circ-\circ$ (second site $\circ$), because the state of the common site (first of the added pair and second of the preceding pair) must coincide. The resulting transfer matrix and its detailed derivation can be found in refs.\cite{Reed1992,Garces2006,Borkovec2006} and it is given by
\begin{equation}
\mathbf{\mathbf{T_{\mathrm{nn}}}}=\left(\begin{array}{cccc}
1 & 0 & z & 0\\
1 & 0 & zu_{\mathrm{nn}} & 0\\
0 & 1 & 0 & zu_{\mathrm{n}}\\
0 & 1 & 0 & zu_{\mathrm{n}}u_{\mathrm{nn}}w
\end{array}\right)\label{eq:TF NNN neighbors}
\end{equation}
where $\epsilon_{\mathrm{nn}}=-\mathrm{k}_{\mathrm{B}}T\ln u_{\mathrm{nn}}$ and $\lambda=-\mathrm{k}_{\mathrm{B}}T\ln w$ represent the next-nearest neighbour and triplet interaction energies respectively. The partition function in this case reads
\begin{equation}
\Xi_{\mathrm{nn}}=\mathbf{v}_{\mathrm{i}}\mathbf{T_{\mathrm{nn}}}^{N}\mathbf{v_{\mathrm{f}}^{T}}
\label{eq:SB partition function}
\end{equation}
with $\mathbf{v}_{\mathrm{i}}=\left(1,0,0,0\right)$ and $\mathbf{v}_{\mathrm{f}}=\left(1,1,1,1\right)$. 

\subsection{Coupling binding and conformations}

Let us use these previous results in order to build the full model which do includes configurations. The conformational state of the FJC is given by the set of $N-2$ angles between consecutive bonds $\gamma_{i}\left(i=3,...,N\right)$. One has to add to the chain a site attached to a bond which forms an angle $\gamma_{i}$ with the previous bonds. The constant pH partition function for a frozen conformational state is given by
\begin{equation}
\Xi\left(\left\{ \gamma_{i}\right\} \right)=\mathbf{v}_{\mathrm{i}}\mathbf{T}_{1}\mathbf{T}_{2}\mathbf{T}_{3}\left(\gamma_{3}\right)\mathbf{T}_{4}\left(\gamma_{4}\right)\cdots\mathbf{T}_{\mathrm{N}}\left(\gamma_{\mathrm{\,N}}\right)\mathbf{v_{\mathrm{f}}^{T}}
\label{eq:frozen angles partition function}
\end{equation}
with
\begin{equation}
\mathbf{\mathbf{T_{\mathrm{i}}}}=\left(\begin{array}{cccc}
1 & 0 & z & 0\\
1 & 0 & zu_{\mathrm{nn}}\left(\gamma_{i}\right) & 0\\
0 & 1 & 0 & zu_{\mathrm{n}}\\
0 & 1 & 0 & zu_{\mathrm{n}}u_{\mathrm{nn}}\left(\gamma_{i}\right)
\end{array}\right)\label{eq:TF angles}
\end{equation}
where the triplet interaction energy $\lambda=0$ ($w=1$) because three-body interactions are not considered at the microscopic level. Be aware that a little trick has been implemented in order to apply the Transfer Matrix Method. Since in this model we do not have next
nearest neighbour interactions neither any angle $\gamma_{i}$ until $N=3$, matrices $\mathbf{T}_{1}$ and $\mathbf{T}_{2}$ are strictly
undefined since no angle exists to determine the value of $u_{\mathrm{nn}}\left(\gamma_{i}\right)$. Nevertheless, when multiplying the initial vector with this two matrices we get that $\mathbf{v}_{\mathrm{i}}\mathbf{T}_{1}\mathbf{T}_{2}=\left(1,z,z,z^{2}u_{\mathrm{n}}\right)$ so the $u_{\mathrm{nn}}\left(\gamma_{i}\right)$ term does not contribute.

The constant pH partition function for all the conformational states is obtained by integrating over the angle variables\cite{Garces2006}, taking into account that the angles are not equiprobable but their cosine 
\begin{equation}
\Xi=\frac{1}{2^{N-2}}\int_{-1}^1\cdots\int_{-1}^1\Xi\left(\left\{ \gamma_{i}\right\} \right)d\cos\gamma_{\,3}\cdots d\cos\gamma_{\mathrm{\,N}}
\end{equation}
Considering that the $N-2$ angles over which we have to average behave identically, the integrals can be individually evaluated and a unique matrix $\mathbf{\mathbf{\tilde{T}_{\mathrm{nn}}}}$ is needed to describe the process. Additionally, as $\mathbf{v}_{\mathrm{i}}\mathbf{T}_{1}\mathbf{T}_{2}=\mathbf{v}_{\mathrm{i}}\mathbf{\mathbf{\tilde{T}_{\mathrm{nn}}}}\mathbf{\mathbf{\tilde{T}_{\mathrm{nn}}}}=\left(1,z,z,z^{2}u_{\mathrm{n}}\right)$
the partition function reads
\begin{equation}
\Xi=\mathbf{v}_{\mathrm{i}}\mathbf{\tilde{T}_{\mathrm{nn}}}^{N}\mathbf{v_{\mathrm{f}}^{T}}
\label{eq:SBRIS NNNN model}
\end{equation}
where
\begin{equation}
\mathbf{\mathbf{\tilde{T}_{\mathrm{nn}}}}=\left(\begin{array}{cccc}
1 & 0 & z & 0\\
1 & 0 & z\tilde{u}_{\mathrm{nn}} & 0\\
0 & 1 & 0 & z\tilde{u}_{\mathrm{n}}\\
0 & 1 & 0 & z\tilde{u}_{\mathrm{n}}\tilde{u}_{\mathrm{nn}}
\end{array}\right)\label{eq: T NNNN model}
\end{equation}
Trivially $\tilde{u}_{\mathrm{n}}=u_{\mathrm{n}}$ as the bonds are rigid and 
\begin{equation}
\tilde{u}_{\mathrm{nn}}=\frac{1}{2}\int_{0}^{\pi}\exp\left(-\beta\epsilon_{\mathrm{nn}}\left(\gamma\right)\right)\sin\left(\gamma\right)d\gamma\label{eq:contracted u NNNN model}
\end{equation}

\section{\label{FiniteChain}Finite Chain}

Let us study now two separate cases of this model: the finite and infinite chain cases. In this section we will start solving the model for the finite chain case where the end effects play a relevant role when calculating the PE's degree of ionization and end to end distance. As we will see, these properties will have an analytical solution that can be solved numerically thanks to the Transfer Matrix Method. For the calculations, parameters are set to $\mathrm{p}K=9.0$ and $b=0.25\,\mathrm{nm}$, similar to those chosen for the polyethyleneimines.\cite{Garces2014}

\subsection{Ionization degree}

The partition function contains all the information about the PE properties. In particular, the degree of protonation $\theta$ (see Appendix Eq.\,\ref{eq:theta NNNN model}) stays
\begin{equation}
\theta=\sum_{s}\theta(s)p(s)=\sum_{s}\left(\frac{\sum_{i}s_{i}}{N}\right)\frac{e^{-\beta F\left(s\right)}}{\Xi}=\frac{1}{N}\frac{\partial\ln\Xi}{\partial\ln z}
\label{eq_theta_finite}
\end{equation}
where the fact that all $\mu_{i}=\mu$ has been taken into account. As depicted in Fig.\,\ref{Fig_FiniteCase}a, the protonation curve at constant ionic strength varies quite a lot for the cases where $N$ is small due to end effects, which cause the PE to be more protonated. This happens because the sites at the extremes suffer less electrostatic repulsion since they do not have neighbours to interact with on both sides, unlike the central sites. Nonetheless, as the length of the PE increases ($N\gtrsim30$), the convergence to the infinite case is almost immediate.

\subsection{End to end distance}

Bond distances are fixed in this model $b_i=b$. Therefore, one can express the end to end distance as
\begin{equation}
\left\langle r^{2}\right\rangle =\left\langle \left(\sum_{i}\overrightarrow{b_{i}}\right)^{2}\right\rangle= \sum_{i}b^{2}+\sum_{i\neq j}\left\langle \overrightarrow{b_{i}}\cdot\overrightarrow{b_{j}}\right\rangle 
\label{eq.r2_bondfixed}
\end{equation}
\begin{figure}[t!]
\includegraphics[scale=0.2,trim= 0.5in 0in 0in 0in]{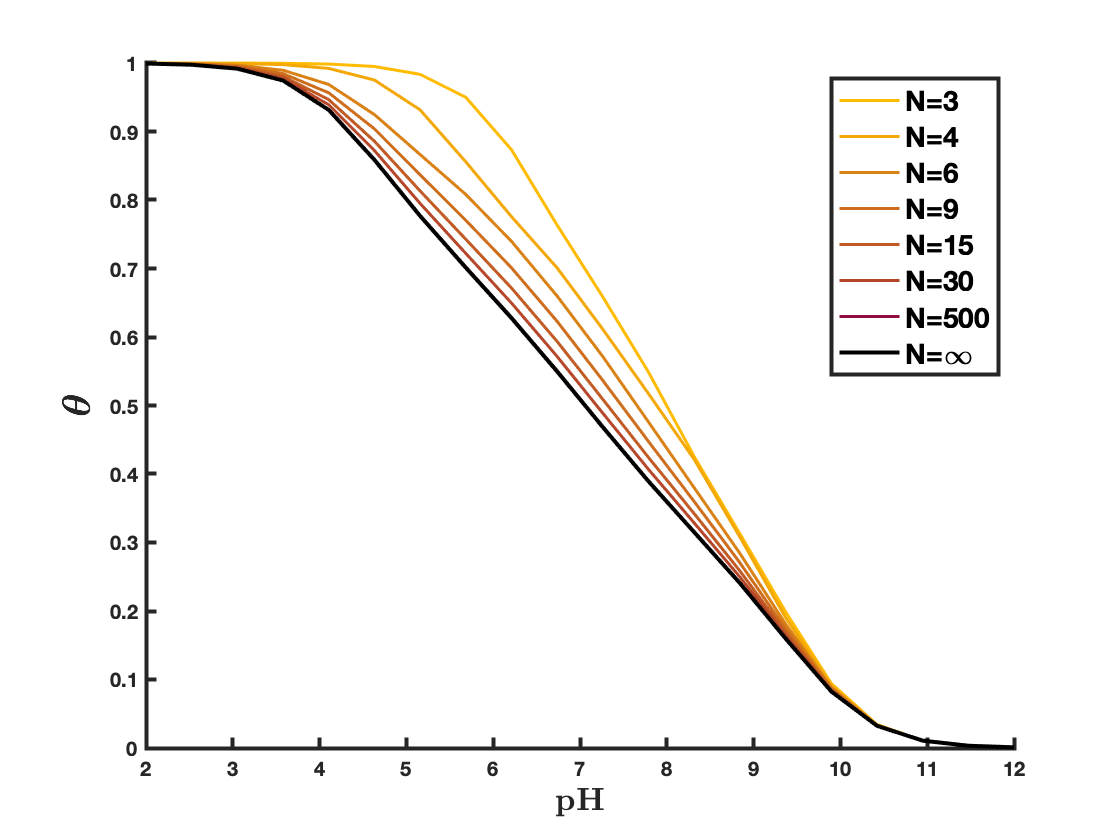}
\begin{center}
    (a)
\end{center}
\includegraphics[scale=0.2,trim= 0.5in 0in 0in 0in]{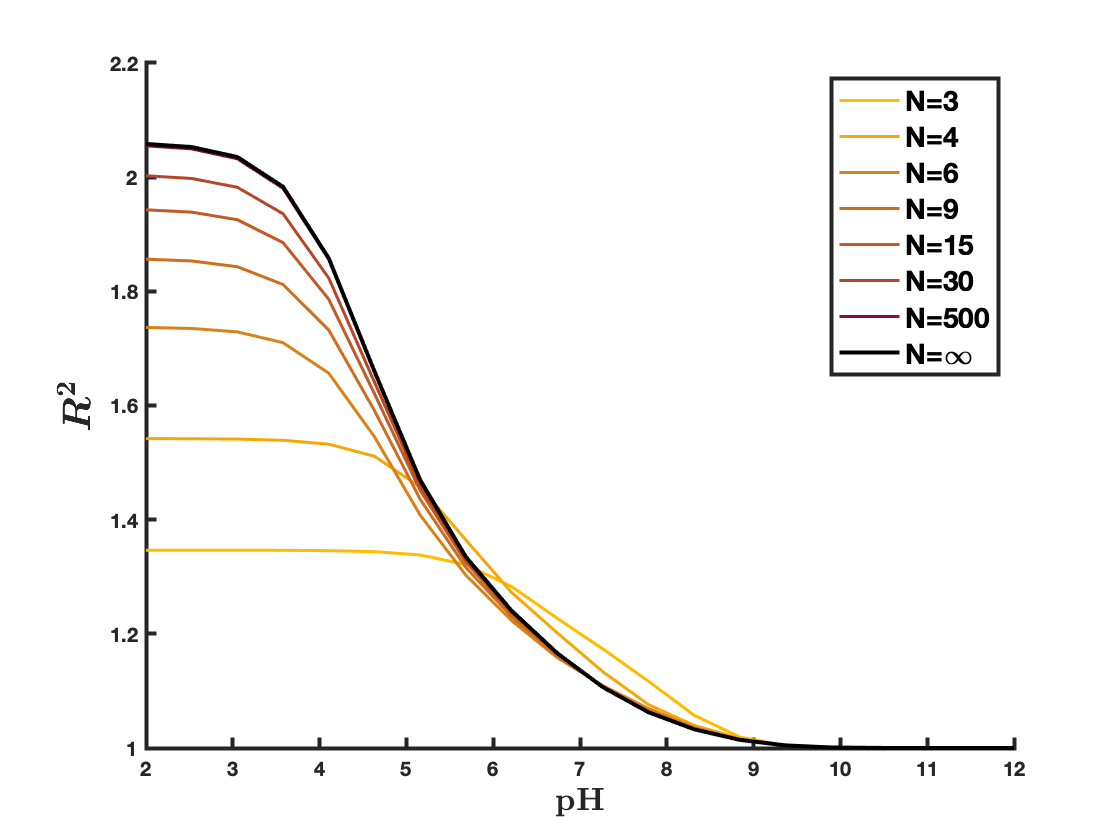}
\begin{center}
    (b)
\end{center}
\caption{(a) Ionization degree $\theta$ and (b) normalized squared end to end distance $R^2={\left\langle r^{2}\right\rangle}/{(N-1)b^{2}}$ for the finite chain with $N$ ranging from $3$ to $500$ interacting sites. The limit case $N\rightarrow\infty$ is also included. Parameters of the model are set to $\mathrm{p}K=9.0$, $I=\mathrm{0.01M}$ and $b=0.25\,\mathrm{nm}$.}
\label{Fig_FiniteCase}
\end{figure}
Now, regarding that the angle sets $\gamma=\left\{\gamma_{i}\right\}$ and $\phi=\left\{ \phi_{i}\right\}$ are independent, and that all $\phi_{i}$ are uncorrelated, the mean dot product $\left\langle \overrightarrow{b_{i}}\cdot\overrightarrow{b_{j}}\right\rangle$ reads
\begin{multline}
\left\langle \overrightarrow{b_{i}}\cdot\overrightarrow{b_{j}}\right\rangle =\left\langle \left(-\cos\gamma_{i+2}\right)\cdots\,(-\cos\gamma_{j+1})\right\rangle=\\=\frac{1}{\Xi}\mathbf{v}_{\mathrm{i}}\left[\mathbf{\tilde{T}_{\mathrm{nn}}}^{(i+1)}\mathbf{\tilde{T}_{\mathrm{cos}}}^{(j-i)}\mathbf{\tilde{T}_{\mathrm{nn}}}^{(N-1-j)}\right]\mathbf{v_{\mathrm{f}}^{T}}
\label{cosineproduct}
\end{multline}
where the new transfer matrix $\mathbf{\tilde{T}_{\mathrm{cos}}}$ is defined as
\begin{equation}
\mathbf{\tilde{T}_{\mathrm{cos}}=}\left(\begin{array}{cccc}
0 & 0 & 0 & 0\\
0 & 0 & z\left\langle \cos\gamma\right\rangle\tilde{u}_{\mathrm{nn}} & 0\\
0 & 0 & 0 & 0\\
0 & 0 & 0 & z\tilde{u}_{\mathrm{n}}\left\langle \cos\gamma\right\rangle\tilde{u}_{\mathrm{nn}}
\end{array}\right)
\end{equation}
and $\left\langle \cos\gamma\right\rangle$ is the average cosine between two next-neighbouring charged sites regardless of whether the middle site is charged or not
\begin{equation}
\left\langle \cos\gamma\right\rangle\tilde{u}_{\mathrm{nn}}=-\frac{1}{2}\int_{-1}^{1}\cos\gamma e^{-\beta\epsilon_{\mathrm{i,i+2}}\left(r\left(\gamma\right)\right)}d\cos\gamma
\end{equation}
Then, replacing Eq.\ref{cosineproduct} into Eq.\ref{eq.r2_bondfixed} the squared end to end distance normalized to the FJC reads
\begin{equation}
\frac{\left\langle r^{2}\right\rangle }{(N-1)b^{2}}=1+\frac{2}{N-1}\sum_{i<j}\frac{1}{\Xi}\mathbf{v}_{\mathrm{i}}\left[\mathbf{\tilde{T}_{\mathrm{nn}}}^{(i+1)}\mathbf{\tilde{T}_{\mathrm{cos}}}^{(j-i)}\mathbf{\tilde{T}_{\mathrm{nn}}}^{(N-1-j)}\right]\mathbf{v_{\mathrm{f}}^{T}}
\label{e2e_exact}
\end{equation}
The complete derivation can be found in the Appendix (Eqs.\,\ref{eq.r2_bondfixed_appendix}-\ref{e2e_appendix}).

Analyzing the results of the normalized end to end distance, it is observed from Fig.\,\ref{Fig_FiniteCase}b that the larger the PE chain is the more extended it gets in almost the whole range of pH values. However, for the extreme short chain cases $N=3$ and $N=4$, end effects cause them to swell more than the rest in the intermediate pH range. It is worth to mention that the normalized end to end distance also tends to the limit $N\rightarrow\infty$ chain, but the convergence is much slower than in the degree of ionization. Calculations with $N\approx500$ are necessary to achieve fairly matching results.

\section{\label{InfiniteChain}Infinite Chain}

Once the finite chain case has been solved, these results can be used to study the particular case where $N\rightarrow\infty$. Here one will observe that relatively compact expressions for the ionization degree and the end to end distance can be obtained from the transfer matrix method itself.

So that the text is easy to follow we define the diagonalization $\mathbf{\tilde{T}_{\mathrm{nn}}}=\mathbf{P}\Lambda_\mathbf{nn}\mathbf{P^{-1}}$, matrix $\mathbf{E}$ and vectors $\mathbf{a^{T}}$ and $\mathbf{b}$ in section \ref{subsec_app_definitions} of the Appendix (Eqs. \ref{prod_tcos_appendix}-\ref{partitionfunc_infinity}). Using this definitions and substituting into Eq.\,\ref{eq:SBRIS NNNN model} the partition function in the limit $N\rightarrow\infty$ can be expressed as
\begin{equation}
\Xi=\mathbf{v}_{\mathrm{i}}\mathbf{\tilde{T}_{\mathrm{nn}}}^{N}\mathbf{v_{\mathrm{f}}^{T}}=\lambda_{max}^{N}\mathbf{v}_{\mathrm{i}}\mathbf{a^{T}b}\mathbf{v_{\mathrm{f}}^{T}}
\end{equation}

\subsection{Ionization degree}

\begin{figure}[t]
\includegraphics[scale=0.2,trim= 0.5in 0in 0in 0in]{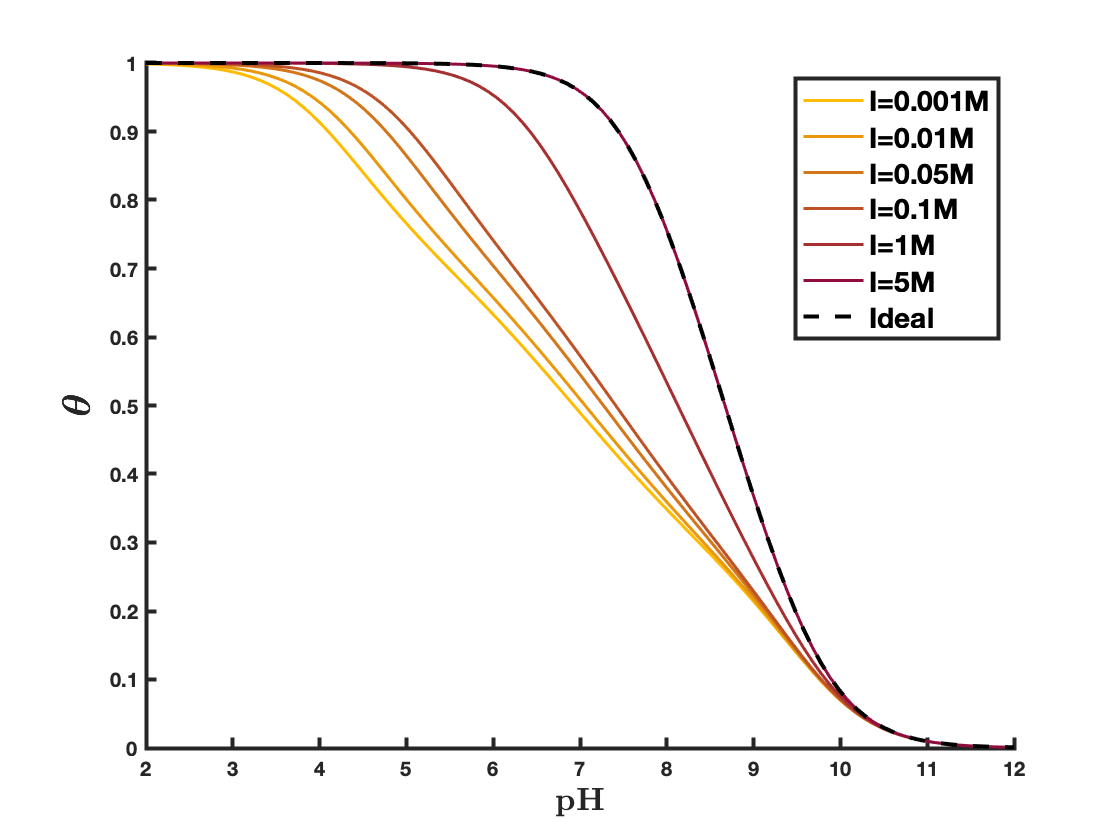}
\begin{center}
    (a)
\end{center}
\includegraphics[scale=0.2,trim= 0.5in 0in 0in 0in]{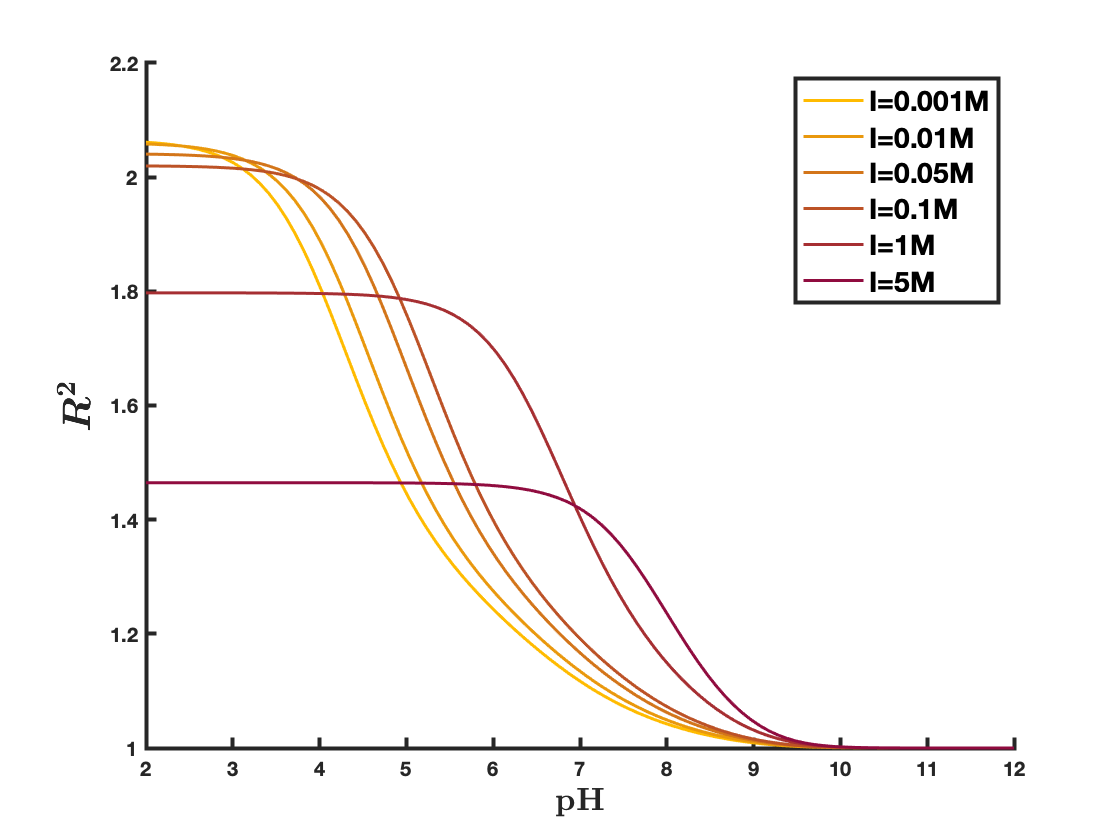}
\begin{center}
    (b)
\end{center}
\caption{(a) Ionization degree $\theta$ and (b) normalized squared end to end distance $R^2={\left\langle r^{2}\right\rangle}/{(N-1)b^{2}}$ for the infinite chain at different ionic strengths ranging from $I=\mathrm{0.001M}$ to $I=\mathrm{5M}$. Model parameters are set to $\mathrm{p}K=9.0$ and $b=0.25\,\mathrm{nm}$.}
\label{Fig_InfiniteCase}
\end{figure}

For the infinite chain case one can assume that all the sites interact in the same way because the end effects are negligible. Therefore, the degree of ionization $\theta$ of the whole PE is equivalent to that of the central site of the chain $\theta_{c}$ and it can be calculated (see Appendix Eq.\,\ref{eq:theta_inf_app}) as follows
\begin{gather}
\nonumber\theta=\theta_{c}=\frac{\partial\ln\Xi}{\partial\ln z_{c}}=\frac{1}{\lambda_{max}}\mathbf{b}\frac{\mathbf{\partial\tilde{T}_{\mathrm{nn}}}}{\partial\ln z_{c}}\mathbf{a^{T}}=\\=\frac{1}{\lambda_{max}}\left[a_{3}\left(b_{1}z+b_{2}z\tilde{u}_{\mathrm{nn}}\right)+a_{4}\left(b_{3}z\tilde{u}_{\mathrm{n}}+b_{4}z\tilde{u}_{\mathrm{n}}\tilde{u}_{\mathrm{nn}}\right)\right]
\end{gather}
As it can be seen from Fig.\,\ref{Fig_InfiniteCase}a the protonation curves for each ionic strength are shifted towards the region of lower pH values compared to the ideal Henderson-Hasselbalch curve. This happens due to the electrostatic repulsion among charged PE sites. The ionic strength shields this effect as it increases, thus at $I=\mathrm{5M}$ we recover the ideal behaviour.
\subsection{Interaction functions}
Following a procedure analogous to that of the preceding section, one can also find the interaction functions that tell us the fraction of two nearest neighbouring sites (Eq.\,\ref{NNpairs}), two next-nearest neighbouring sites with the middle site uncharged (Eq.\,\ref{NNNpairs}) or three consecutive sites (Eq.\,\ref{triplets}) charged at the same time. The complete derivation is provided in the Appendix (Eqs.\,\ref{interaction functions}).
\begin{subequations}
\begin{gather}
\varphi_{\bullet\bullet\circ}=\left(\varphi_{\bullet\bullet\circ}\right)_{c}=\frac{1}{\lambda_{max}}\left[a_{4}b_{3}z\tilde{u}_{\mathrm{n}}\right]
\label{NNpairs}\\
\varphi_{\bullet\circ\bullet}=\left(\varphi_{\bullet\circ\bullet}\right)_{c}=\frac{1}{\lambda_{max}}\left[a_{3}b_{2}z\tilde{u}_{\mathrm{nn}}\right]
\label{NNNpairs}\\
\xi_{\bullet\bullet\bullet}=\left(\xi_{\bullet\bullet\bullet}\right)_{c}=\frac{1}{\lambda_{max}}\left[a_{4}b_{4}z\tilde{u}_{\mathrm{n}}\tilde{u}_{\mathrm{nn}}\right]
\label{triplets}
\end{gather}
\label{interactionfunc}
\end{subequations}

\subsection{End to end distance}

Recovering Eq.\,\ref{cosineproduct} for the finite case, and considering the diagonalization of $\mathbf{\tilde{T}_{\mathrm{nn}}}$, the cosine product average in the limit where $N\rightarrow\infty$ can be expressed as
\begin{equation}
\left\langle \left(-\cos\gamma_{i+2}\right)\cdots\,(-\cos\gamma_{j+1})\right\rangle =\frac{1}{\lambda_{max}^{\left(j-i\right)}}\left[\mathbf{b}\mathbf{\tilde{T}_{\mathrm{cos}}}^{(j-i)}\mathbf{a^{T}}\right]
\end{equation}
where one can observe that the interaction only depends on the difference between $j$ and $i$ as expected for the infinite case. Therefore, reformulating the indexes according to $k=j-i$, and defining the terms $A_{k}=\left[\mathbf{b}\mathbf{\tilde{T}_{\mathrm{cos}}}^{k}\mathbf{a^{T}}\right]/\lambda_{max}^{k}$, Eq.\,\ref{e2e_exact} takes the following form
\begin{equation}
\frac{\left\langle r^{2}\right\rangle }{(N-1)b^{2}}=1+2\sum_{k=1}^{N-2}A_{k}-\frac{2}{N-1}\sum_{k=1}^{N-2}kA_{k}
\end{equation}
Analyzing the sums and applying the relations from (Eqs.\,\ref{interactionfunc}) we get
\begin{equation}
\frac{\left\langle r^{2}\right\rangle }{(N-1)b^{2}}=1+2\left\langle \cos\gamma\right\rangle \left[\varphi_{\bullet\circ\bullet}+\frac{1}{1-\alpha}\xi_{\bullet\bullet\bullet}\right]
\end{equation}
where $\alpha=\left(z\tilde{u}_{\mathrm{n}}\left\langle \cos\gamma\right\rangle\tilde{u}_{\mathrm{nn}}\right)/\lambda_{max}$. All the steps are clearly detailed in the Appendix (Eqs.\,\ref{cosineproduct_infinity}-\ref{e2e_infinity_appendix}).

Regarding Fig.\,\ref{Fig_InfiniteCase}b we observe that the end to end distance increases considerably when we diminish the ionic strength at low pH values. This is again a consequence of the loss of shielding that the charged PE suffers and that causes it to stretch due to repulsion. On the other hand, in the region of high pH values with the discharged PE we recover the behaviour of a FJC as expected.

\section{Conclusions}
In general, the effect of charge regulation coupled to conformations in weak PEs is a complex task to solve analytically due to the emerging correlations among sites and bonds. However, in the model considered, despite being very flexible, the transfer matrix method is applicable thanks to the independence of the matrices when performing the conformational integration. This fact allows us to obtain a completely exact model that can be useful to study experimental results in which short range conditions are applicable.\cite{Smits1993,Borkovec1996,Garces2006,Garces2014} In relation with the ionization curves, short chains are more charged than long ones comparing at the same pH levels, due to end effects. Convergence of the ionization degree $\theta$ to the infinite chain case is really fast ($N\gtrsim30$). As for the end to end distance, we can assure that long PEs become more extended at almost all pH values, except for the cases $N=3$ and $N=4$ for which the trend does not apply in the intermediate pH range. In this case the convergence to the infinite chain ($N\rightarrow\infty$) is much slower and values of $N\approx500$ are needed to find matching results.
\\
\\

\section{Aknowledgements}
P.M.B., S.M. and F.M. acknowledge the financial support from Generalitat de Catalunya (Grant 2021SGR00350). S.M. and F.M. acknowledge Spanish Structures of Excellence María de Maeztu program through Grant CEX2021-001202-M. J.L.G. also acknowledges the Spanish Ministry of Science and Innovation (Project PID2022-140312NB-C21). J.O. acknowledges the financial support of the FPU grant from Spainsh Ministry of Universities. P.M.B. acknowledges the funding from the European Union\textquoteright s Horizon Europe research and innovation programme under the Marie Sklodowska-Curie grant agreement No 101062456.

\section{References}

\nocite{*}
\bibliographystyle{ieeetr}
\bibliography{aipsamp}

\begin{thebibliography}{10}

\bibitem{Blanco2019b}
P.~M. Blanco, S.~Madurga, C.~F. Narambuena, F.~Mas, and J.~L. Garc{\'{e}}s, ``{Role of Charge Regulation and Fluctuations in the Conformational and Mechanical Properties of Weak Flexible Polyelectrolytes},'' {\em Polymers}, vol.~11, p.~1962, 2019.

\bibitem{Garces2017}
J.~L. Garc{\'{e}}s, S.~Madurga, C.~Rey-Castro, and F.~Mas, ``{Dealing with long-range interactions in the determination of polyelectrolyte ionization properties. Extension of the transfer matrix formalism to the full range of ionic strengths},'' {\em Journal of Polymer Science, Part B: Polymer Physics}, vol.~55, no.~3, pp.~275--284, 2017.

\bibitem{Hartig2007}
S.~M. Hartig, R.~R. Greene, M.~M. Dikov, A.~Prokop, and J.~M. Davidson, ``{Multifunctional nanoparticulate polyelectrolyte complexes},'' {\em Pharmaceutical Research}, vol.~24, no.~12, pp.~2353--2369, 2007.

\bibitem{Trefalt2016}
G.~Trefalt, S.~H. Behrens, and M.~Borkovec, ``{Charge Regulation in the Electrical Double Layer: Ion Adsorption and Surface Interactions},'' {\em Langmuir}, vol.~32, no.~2, pp.~380--400, 2016.

\bibitem{Li2016}
Y.~Li, T.~Zhao, C.~Wang, Z.~Lin, G.~Huang, B.~D. Sumer, and J.~Gao, ``{Molecular basis of cooperativity in pH-triggered supramolecular self-assembly},'' {\em Nature Communications}, vol.~7, pp.~1--9, 2016.

\bibitem{Blanco2023}
P.~Blanco, C.~Narambuena, S.~Madurga, F.~Mas, and J.~Garcés, ``Unusual aspects of charge regulation in flexible weak polyelectrolytes,'' {\em Polymers}, vol.~15, p.~2680, 2023.

\bibitem{Marcus1954}
R.~A. Marcus, ``Titration of polyelectrolytes at higher ionic strengths,'' {\em Journal of Physical Chemistry}, vol.~58, no.~8, p.~621–623, 1954.

\bibitem{Reed1992}
C.~E. Reed and W.~F. Reed, ``{Monte Carlo study of titration of linear polyelectrolytes},'' {\em Journal of Chemical Physics}, vol.~96, no.~2, p.~1609, 1992.

\bibitem{Smits1993}
R.~Smits, G.~Koper, and M.~Mandel, ``The influence of nearest-and next-nearest-neighbor interactions on the potentiometric titration of linear poly (ethylenimine),'' {\em Journal of Physical Chemistry}, vol.~97, no.~21, pp.~5745--5751, 1993.

\bibitem{Flory1969}
P.~J. Flory, {\em Statistical Mechanics of Chain Molecules}.
\newblock Wiley, 1969.

\bibitem{Chandler1987}
D.~Chandler, {\em Introduction to Modern Statistical Mechanics}.
\newblock Oxford University Press, 1987.

\bibitem{Koper1996}
G.~J. Koper and M.~Borkovec, ``{Exact affinity distributions for linear polyampholytes and polyelectrolytes},'' {\em Journal of Chemical Physics}, vol.~104, no.~11, pp.~4204--4213, 1996.

\bibitem{Borkovec1996}
G.~J. M.~K. M.~Borkovec, ``Ising models and acid-base properties of weak polyelectrolytes,'' {\em Berichte der Bunsengesellschaft für physikalische Chemie}, vol.~100, no.~6, pp.~764--769, 1996.

\bibitem{Borkovec2006}
M.~Borkovec, G.~J.~M. Koper, and C.~Piguet, ``{Ion binding to polyelectrolytes},'' {\em Current Opinion in Colloid and Interface Science}, vol.~11, no.~5, pp.~280--289, 2006.

\bibitem{Garces2006}
J.~L. Garcés, G.~J. Koper, and M.~Borkovec, ``{Ionization Equilibria and Conformational Transitions in Polyprotic Molecules and Polyelectrolytes},'' {\em Journal of Physical Chemistry B}, vol.~110, pp.~10937--10950, 2006.

\bibitem{Garces2014}
J.~L. Garc{\'{e}}s, S.~Madurga, and M.~Borkovec, ``{Coupling of conformational and ionization equilibria in linear poly(ethylenimine): A study based on the site binding/rotational isomeric state (SBRIS) model},'' {\em Physical Chemistry Chemical Physics}, vol.~16, no.~10, pp.~4626--4638, 2014.

\end{thebibliography}

\appendix

\begin{widetext}
\section{Finite Chain}

\subsection{\label{Ionization_app_finite}Ionization Degree}
The derivation of the ionization degree expression from the partition function is shown below. The degree of ionization is the average of the degrees of ionization in each protonation state weighted by the probability of being in that particular state $p(s)$. If one rewrites this probability in terms of the Boltzmann factors ($e^{-\beta F(s)}$) and the partition function we have that
\begin{equation}
\theta=\sum_{s}\theta(s)p(s)=\sum_{s}\left(\frac{\sum_{i}s_{i}}{N}\right)\frac{e^{-\beta\left[\mu\sum_{i}s_{i}+\tilde{\epsilon}_{\mathrm{n}}\sum_{i}s_{i}s_{i+1}+\tilde{\epsilon}_{\mathrm{nn}}\sum_{i}s_{i}s_{i+2}\right]}}{\Xi}
\end{equation}
Now notice that the sum of $s_i$ can be replaced by the derivative of the Boltzmann factor with respect to the reduced chemical potential $\mu$, and exchanging the order of the derivative and the summation, one obtains Eq.\,\ref{eq_theta_finite} presented in the main text
\begin{equation}
\theta=\frac{1}{N\Xi}\sum_{s}\frac{\partial}{\partial\left(-\beta\mu\right)}\left(e^{-\beta\left[\mu\sum_{i}s_{i}+\tilde{\epsilon}_{\mathrm{n}}\sum_{i}s_{i}s_{i+1}+\tilde{\epsilon}_{\mathrm{nn}}\sum_{i}s_{i}s_{i+2}\right]}\right)=\frac{1}{N\Xi}\frac{\partial\left[\sum_{s}e^{-\beta F\left(s\right)}\right]}{\partial\left(-\beta\mu\right)}=\frac{1}{N}\frac{\partial\ln\Xi}{\partial\ln z} 
\label{eq:theta NNNN model}
\end{equation}

\subsection{\label{subsec_app_e2e_finite}End to End Distance}

In order to demonstrate the expression found for the end to end distance (Eq.\,\ref{e2e_exact}), we must first consider that the bonds in our PE model are rigid and have length $b$. Assuming this, the expression for the square of the end to end distance becomes
\begin{equation}
\left\langle r^{2}\right\rangle =\left\langle \left(\sum_{i}\overrightarrow{b_{i}}\right)^{2}\right\rangle =\left\langle \left(\sum_{i}\overrightarrow{b_{i}}\right)\left(\sum_{j}\overrightarrow{b_{j}}\right)\right\rangle =\sum_{i}b^{2}+\sum_{i\neq j}\left\langle \overrightarrow{b_{i}}\cdot\overrightarrow{b_{j}}\right\rangle
\label{eq.r2_bondfixed_appendix}
\end{equation}
Then, it will be necessary to express the mean scalar product $\left\langle \overrightarrow{b_{i}}\cdot\overrightarrow{b_{j}}\right\rangle$ of the bonds in an alternative way. For this purpose, the rotation matrices from Flory\cite{Flory1969} can be employed
\begin{equation}
\left\langle \overrightarrow{b_{i}}\cdot\overrightarrow{b_{j}}\right\rangle =b_{i}b_{j}\left\langle \mathbf{R}_{i}\cdots\mathbf{R}_{j-1}\right\rangle _{1,1}=b_{i}b_{j}\left[\left\langle \left\langle \mathbf{R}_{i}\cdots\mathbf{R}_{j-1}\right\rangle _{\phi}\right\rangle _{\beta}\right]_{1,1}=b^{2}\left[\left\langle \left\langle \mathbf{R}_{i}\right\rangle _{\phi_{i}}\cdots\left\langle \mathbf{R}_{j-1}\right\rangle _{\phi_{j-1}}\right\rangle _{\beta}\right]_{1,1}
\end{equation}
where it has been used that the angle sets $\beta=\left\{ \beta_{i}\right\}$ (bond angles) and $\phi=\left\{ \phi_{i}\right\}$ (dihedral angles) are independent, as well as that all $\phi_{i}$ are uncorrelated too. This is aplicable in this case since only next-nearest neighbour interactions are considered, and no restrictions appear on $\phi_{i}$ variables. Determining then that the firrst element of the matrix product reads 
\begin{equation}
\left[\left\langle \left\langle \mathbf{R}_{i}\right\rangle _{\phi_{i}}\cdots\left\langle \mathbf{R}_{j-1}\right\rangle _{\phi_{j-1}}\right\rangle _{\beta}\right]_{1,1}=\left[\left\langle \left(\begin{array}{ccc}
\cos\beta_{i} & \sin\beta_{i} & 0\\
0 & 0 & 0\\
0 & 0 & 0
\end{array}\right)\cdots\left(\begin{array}{ccc}
\cos\beta_{j-1} & \sin\beta_{j-1} & 0\\
0 & 0 & 0\\
0 & 0 & 0
\end{array}\right)\right\rangle _{\beta}\right]_{1,1}=\left\langle \cos\beta_{i}\cdots\cos\beta_{j-1}\right\rangle _{\beta}
\end{equation}
and taking into account that Flory uses the complementary angle $\beta_{i}$ instead of the bond angle $\gamma_{i}$, a sign change must appear in the cosine and the square end to end distance is
\begin{equation}
\left\langle r^{2}\right\rangle =\sum_{i}b^{2}+\sum_{i\neq j}b^{2}\left\langle \cos\beta_{i+2}\cdots\cos\beta_{j+1}\right\rangle =\sum_{i}b^{2}+\sum_{i\neq j}b^{2}\left\langle (-\cos\gamma_{i+2})\cdots\,(-\cos\gamma_{j+1})\right\rangle 
\label{eq:r2_productcosine_appendix}
\end{equation}
Now considering that the average of the cosines can be expressed as the integral of the product of the cosines multiplied by the probability of being in a certain state with an angle set $\gamma=\left\{\gamma_{i}\right\}$, and that all variables can be integrated separately.
\begin{gather}
\nonumber\left\langle \left(-\cos\gamma_{i+2}\right)\cdots\,(-\cos\gamma_{j+1})\right\rangle=\frac{1}{2^{N-2}}\int_{-1}^1\cdots\int_{-1}^1\frac{\Xi\left(\left\{ \gamma_{i}\right\} \right)}{\Xi}\,(-\cos\gamma_{i+2})\cdots\,(-\cos\gamma_{j+1})\,\frac{d\cos\gamma_{3}}{2}\cdots\,\frac{d\cos\gamma_{N}}{2}=\\
\nonumber\\
=\frac{1}{\Xi}\mathbf{v}_{\mathrm{i}}\left[\mathbf{\tilde{T}_{\mathrm{nn}}}^{(i+1)}\mathbf{\tilde{T}_{\mathrm{cos}}}^{(j-i)}\mathbf{\tilde{T}_{\mathrm{nn}}}^{(N-1-j)}\right]\mathbf{v_{\mathrm{f}}^{T}}
\label{productcosine_appendix}
\end{gather}
where the matrix $\mathbf{\tilde{T}_{\mathrm{cos}}}$ reads 
\begin{equation}
\mathbf{\tilde{T}_{\mathrm{cos}}=}\left(\begin{array}{cccc}
0 & 0 & 0 & 0\\
0 & 0 & z\left\langle \cos\gamma\right\rangle\tilde{u}_{\mathrm{nn}} & 0\\
0 & 0 & 0 & 0\\
0 & 0 & 0 & z\tilde{u}_{\mathrm{n}}\left\langle \cos\gamma\right\rangle\tilde{u}_{\mathrm{nn}}
\end{array}\right)\qquad with\qquad\left\langle \cos\gamma\right\rangle\tilde{u}_{\mathrm{nn}}=-\frac{1}{2}\int_{-1}^{1}\cos\gamma e^{-\beta\epsilon_{\mathrm{i,i+2}}\left(r\left(\gamma\right)\right)}d\cos\gamma
\label{eq:tcos_appendix}
\end{equation}
Therefore, substituting this result into Eq. \ref{eq:r2_productcosine_appendix} the square end to end distance normalized to the FJC reads
\begin{equation}
\frac{\left\langle r^{2}\right\rangle }{(N-1)b^{2}}=1+\frac{2}{N-1}\sum_{i<j}\frac{1}{\Xi}\mathbf{v}_{\mathrm{i}}\left[\mathbf{\tilde{T}_{\mathrm{nn}}}^{(i+1)}\mathbf{\tilde{T}_{\mathrm{cos}}}^{(j-i)}\mathbf{\tilde{T}_{\mathrm{nn}}}^{(N-1-j)}\right]\mathbf{v_{\mathrm{f}}^{T}}
\label{e2e_appendix}
\end{equation}

\section{Infinite Chain}

In order to study the PE in the $N\rightarrow\infty$ let us first define some new matrices and vectors that will be very useful to describe the system.

\subsection{\label{subsec_app_definitions}Definitions}
First we define the product $\mathbf{\tilde{T}_{\mathrm{cos}}^{(j-i)}}$, which is a matrix that consists of only one element and reads
\begin{equation}
\mathbf{\tilde{T}_{\mathrm{cos}}^{(j-i)}}=\left(\begin{array}{cccc}
0 & 0 & 0 & 0\\
0 & 0 & 0 & 0\\
0 & 0 & 0 & 0\\
0 & 0 & 0 & \left[z\tilde{u}_{\mathrm{n}}\left\langle \cos\gamma\right\rangle\tilde{u}_{\mathrm{nn}}\right]^{j-i}
\end{array}\right)
\label{prod_tcos_appendix}
\end{equation}
when $\:j\neq i+1$. It is also vital to determine the diagonalization of the $\mathbf{\tilde{T}_{\mathrm{nn}}}$ matrix 
\begin{equation}
\mathbf{\tilde{T}_{\mathrm{nn}}}=\mathbf{P}\Lambda_\mathbf{nn}\mathbf{P^{-1}}\qquad where\qquad\Lambda_\mathbf{nn}=\left(\begin{array}{cccc}
\lambda_{1,1} & 0 & 0 & 0\\
0 & \lambda_{2,2} & 0 & 0\\
0 & 0 & \lambda_{3,3} & 0\\
0 & 0 & 0 & \lambda_{4,4}
\end{array}\right)
\end{equation}
Additionally, as $\Lambda_\mathbf{nn}$ is a diagonal matrix, when it is raised to a value $s\rightarrow\infty$ one can perform the following manipulation and obtain matrix $\mathbf{E}$, an array of zeros except for a $1$ in the position of $\lambda_{max}=max(\lambda_{r,r})$ 
\begin{equation}
\mathbf{E}=\lim_{s\rightarrow\infty}\left(\frac{\Lambda_\mathbf{nn}}{\lambda_{max}}\right)=\left(\begin{array}{cccc}
0 & 0 & 0 & 0\\
0 & 0 & 0 & 0\\
0 & 0 & 1 & 0\\
0 & 0 & 0 & 0
\end{array}\right)
\end{equation}
where we have assumed that eigenvalue $\lambda_{max}=\lambda_{3,3}$ as an example. This $\mathbf{E}$ matrix allows us to express the product 
\begin{equation}
\mathbf{P}\Lambda_\mathbf{nn}^{s}\mathbf{P^{-1}}=\lambda_{max}^{s}\mathbf{P}\mathbf{E}^{s}\mathbf{P^{-1}}=\lambda_{max}^{s}\mathbf{P}\mathbf{E}\mathbf{P^{-1}}=\lambda_{max}^{s}\mathbf{a^{T}b}
\label{product_PEP}
\end{equation}
where vector $\mathbf{a^{T}}=\left(\mathbf{P}_{1,r},\mathbf{P}_{2,r},\mathbf{P}_{3,r},\mathbf{P}_{4,r}\right)$ is the $r$-th column of $\mathbf{P}$ and vector $\mathbf{b}=\left(\mathbf{P^{-1}}_{r,1},\mathbf{P^{-1}}_{r,2},\mathbf{P^{-1}}_{r,3},\mathbf{P^{-1}}_{r,4}\right)$ is the $r$-th row of $\mathbf{P^{-1}}$. Then, regarding Eq.\,\ref{product_PEP} the partition function in the environment where $N\rightarrow\infty$ reads
\begin{equation}
\Xi=\mathbf{v}_{\mathrm{i}}\mathbf{\tilde{T}_{\mathrm{nn}}}^{N}\mathbf{v_{\mathrm{f}}^{T}}=\mathbf{v}_{\mathrm{i}}\mathbf{P}\Lambda_{nn}^{N}\mathbf{P^{-1}}\mathbf{v_{\mathrm{f}}^{T}}=\lambda_{max}^{N}\mathbf{v}_{\mathrm{i}}\mathbf{P}\mathbf{E}^{N}\mathbf{P^{-1}}\mathbf{v_{\mathrm{f}}^{T}}=\lambda_{max}^{N}\mathbf{v}_{\mathrm{i}}\mathbf{a^{T}b}\mathbf{v_{\mathrm{f}}^{T}}
\label{partitionfunc_infinity}
\end{equation}

\subsection{Ionization Degree}
Without loss of generality in the infinite chain case we are dealing with one can consider that $N=2m+1$ sites with $m\rightarrow\infty$. Using this trick and taking into account that the protonation properties of the chain will be the same everywhere (since no end-effects are present), we can choose the central site to study them. Therefore, using the ionization degree expression from Eq.\,\ref{eq:theta NNNN model} and the product from Eq.\,\ref{product_PEP} we get that
\begin{gather}
\nonumber\theta=\theta_{c}=\frac{\partial\ln\Xi}{\partial\ln z_{c}}=\frac{1}{\Xi}\frac{\partial\left[\mathbf{v}_{\mathrm{i}}\mathbf{\tilde{T}_{\mathrm{nn}}^{N}}\mathbf{v_{\mathrm{f}}^{T}}\right]}{\partial\ln z_{c}}=\frac{1}{\Xi}\left[\mathbf{v}_{\mathrm{i}}\mathbf{\tilde{T}_{\mathrm{nn}}^{m}}\frac{\mathbf{\partial\tilde{T}_{\mathrm{nn}}}}{\partial\ln z_{c}}\mathbf{\mathbf{\tilde{T}_{\mathrm{nn}}^{m}}v_{\mathrm{f}}^{T}}\right]=\frac{1}{\Xi}\left[\mathbf{v}_{\mathrm{i}}\mathbf{P}\mathbf{E^{m}}\mathbf{P^{-1}}\frac{\mathbf{\partial\tilde{T}_{\mathrm{nn}}}}{\partial\ln z_{c}}\mathbf{P}\mathbf{E^{m}}\mathbf{P^{-1}}\mathbf{v_{\mathrm{f}}^{T}}\right]=\\
=\frac{\lambda_{max}^{2m}\mathbf{v}_{\mathrm{i}}\mathbf{a^{T}b}\frac{\mathbf{\partial\tilde{T}_{\mathrm{nn}}}}{\partial\ln z_{c}}\mathbf{a^{T}b}\mathbf{v_{\mathrm{f}}^{T}}}{\lambda_{max}^{2m+1}\mathbf{v}_{\mathrm{i}}\mathbf{a^{T}b}\mathbf{v_{\mathrm{f}}^{T}}}=\frac{1}{\lambda_{max}}\mathbf{b}\frac{\mathbf{\partial\tilde{T}_{\mathrm{nn}}}}{\partial\ln z_{c}}\mathbf{a^{T}}=\frac{1}{\lambda_{max}}\left[a_{3}\left(b_{1}z+b_{2}z\tilde{u}_{\mathrm{nn}}\right)+a_{4}\left(b_{3}z\tilde{u}_{\mathrm{n}}+b_{4}z\tilde{u}_{\mathrm{n}}\tilde{u}_{\mathrm{nn}}\right)\right]
\label{eq:theta_inf_app}
\end{gather}

\subsection{\label{subsec_app_interactfunc_infinite}Interaction functions}
Let us consider some functions that allow us to know how the sites of the chain interact at different pH values. As the electrostatic interactions in this model can only take place with first or second neighbours we are interested in defining three interaction functions: the fraction of having two consecutive charged sites, of having two charged sites separated by one uncharged, or of having three consecutive sites charged.

In order to obtain these interaction functions it is only necessary to start from its definition: the product among the sites that are considered. If all the sites we want to interact are charged, the product of all $s_i$ will be 1, otherwise it will be zero 0. Be aware that in the first two functions related to pairs (Eqs. \ref{NNpairs_app} and \ref{NNNpairs_app}), as no restriction appears in $s_{c+2}$ and in $s_{c+1}$ respectively, this sites can either be charged or uncharged. Therefore we will have to subtract the fraction of triplets (three consecutive charged sites) to consider only the real pairs with one uncharged site.
\begin{subequations}
\begin{gather}
\nonumber\varphi_{\bullet\bullet\circ}=\left(\varphi_{\bullet\bullet\circ}\right)_{c}=\sum_{s}\left[(s_cs_{c+1})-(s_cs_{c+1}s_{c+2})\right]p(s)=\frac{\partial\ln\Xi}{\partial\ln(\tilde{u}_{\mathrm{n}})_{c}}--\frac{1}{\Xi}\frac{\partial^{2}\Xi}{\partial\ln(\tilde{u}_{\mathrm{n}})_{c}\partial\ln(\tilde{u}_{\mathrm{nn}})_{c}}=\\=\frac{1}{\lambda_{max}}\left[\mathbf{b}\frac{\mathbf{\partial\tilde{T}_{\mathrm{nn}}}}{\partial\ln(\tilde{u}_{\mathrm{n}})_{c}}\mathbf{a^{T}}-\frac{1}{\lambda_{max}}\mathbf{b}\frac{\partial^{2}\mathbf{\tilde{T}_{\mathrm{nn}}}}{\partial\ln(\tilde{u}_{\mathrm{n}})_{c}\partial\ln(\tilde{u}_{\mathrm{nn}})_{c}}\mathbf{a^{T}}\right]=\frac{1}{\lambda_{max}}\left[a_{4}b_{3}z\tilde{u}_{\mathrm{n}}\right]
\label{NNpairs_app}
\end{gather}
\begin{gather}
\nonumber\varphi_{\bullet\circ\bullet}=\left(\varphi_{\bullet\circ\bullet}\right)_{c}=\sum_{s}\left[(s_cs_{c+2})-(s_cs_{c+1}s_{c+2})\right]p(s)=\frac{\partial\ln\Xi}{\partial\ln(\tilde{u}_{\mathrm{nn}})_{c}}-\frac{1}{\Xi}\frac{\partial^{2}\Xi}{\partial\ln(\tilde{u}_{\mathrm{n}})_{c}\partial\ln(\tilde{u}_{\mathrm{nn}})_{c}}=\\=\frac{1}{\lambda_{max}}\left[\mathbf{b}\frac{\mathbf{\partial\tilde{T}_{\mathrm{nn}}}}{\partial\ln(\tilde{u}_{\mathrm{nn}})_{c}}\mathbf{a^{T}}-\frac{1}{\lambda_{max}}\mathbf{b}\frac{\partial^{2}\mathbf{\tilde{T}_{\mathrm{nn}}}}{\partial\ln(\tilde{u}_{\mathrm{n}})_{c}\partial\ln(\tilde{u}_{\mathrm{nn}})_{c}}\mathbf{a^{T}}\right]=\frac{1}{\lambda_{max}}\left[a_{3}b_{2}z\tilde{u}_{\mathrm{nn}}\right]
\label{NNNpairs_app}
\end{gather}
\begin{equation}
\xi_{\bullet\bullet\bullet}=\left(\xi_{\bullet\bullet\bullet}\right)_{c}=\sum_{s}(s_cs_{c+1}s_{c+2})p(s)=\frac{1}{\Xi}\frac{\partial^{2}\Xi}{\partial\ln(\tilde{u}_{\mathrm{n}})_{c}\partial\ln(\tilde{u}_{\mathrm{nn}})_{c}}=\frac{1}{\lambda_{max}}\mathbf{b}\frac{\partial^{2}\mathbf{\tilde{T}_{\mathrm{nn}}}}{\partial\ln(\tilde{u}_{\mathrm{n}})_{c}\partial\ln(\tilde{u}_{\mathrm{nn}})_{c}}\mathbf{a^{T}}=\frac{1}{\lambda_{max}}\left[a_{4}b_{4}z\tilde{u}_{\mathrm{n}}\tilde{u}_{\mathrm{nn}}\right]
\label{trip_app}
\end{equation}
\label{interaction functions}
\end{subequations}

\subsection{\label{subsec_app_e2e_infinite}End to End Distance}
Recovering equation (\ref{productcosine_appendix}) for the finite case, if one considers the diagonalization of $\mathbf{\tilde{T}_{\mathrm{nn}}}$ and the terms $i+1,N-1-j\rightarrow\infty$, the cosine product can be expressed as
\begin{gather}
\nonumber\left\langle \left(-\cos\gamma_{i+2}\right)\cdots\,(-\cos\gamma_{j+1})\right\rangle =\frac{1}{\Xi}\mathbf{v}_{\mathrm{i}}\left[\mathbf{P}\Lambda_\mathbf{nn}^{(i+1)}\mathbf{P^{-1}}\mathbf{\tilde{T}_{\mathrm{cos}}}^{(j-i)}\mathbf{P}\Lambda_\mathbf{nn}^{(N-1-j)}\mathbf{P^{-1}}\right]\mathbf{v_{\mathrm{f}}^{T}}=\\=\frac{\lambda_{max}^{\left(i+1+N-1-j\right)}}{\Xi}\mathbf{v}_{\mathrm{i}}\left[\mathbf{P}\mathbf{E}^{(i+1)}\mathbf{P^{-1}}\mathbf{\tilde{T}_{\mathrm{cos}}}^{(j-i)}\mathbf{P}\mathbf{E}^{(N-1-j)}\mathbf{P^{-1}}\right]\mathbf{v_{\mathrm{f}}^{T}=}\frac{\lambda_{max}^{\left(i+N-j\right)}}{\Xi}\mathbf{v}_{\mathrm{i}}\left[\mathbf{a^{T}b}\mathbf{\tilde{T}_{\mathrm{cos}}}^{(j-i)}\mathbf{a^{T}b}\right]\mathbf{v_{\mathrm{f}}^{T}}
\label{cosineproduct_infinity}
\end{gather}
Regarding the partition function in the $N\rightarrow\infty$ limit (Eq.\,\ref{partitionfunc_infinity}), the cosine product becomes
\begin{equation}
\left\langle \left(-\cos\gamma_{i+2}\right)\cdots\,(-\cos\gamma_{j+1})\right\rangle =\frac{\lambda_{max}^{\left(i+N-j\right)}\mathbf{v}_{\mathrm{i}}\left[\mathbf{a^{T}b}\mathbf{\tilde{T}_{\mathrm{cos}}}^{(j-i)}\mathbf{a^{T}b}\right]\mathbf{v_{\mathrm{f}}^{T}}}{\lambda_{max}^{N}\mathbf{v}_{\mathrm{i}}\mathbf{a^{T}b}\mathbf{v_{\mathrm{f}}^{T}}}=\frac{1}{\lambda_{max}^{\left(j-i\right)}}\left[\mathbf{b}\mathbf{\tilde{T}_{\mathrm{cos}}}^{(j-i)}\mathbf{a^{T}}\right]
\label{cosineproduct_infinity2}
\end{equation}
Now replacing Eq.\,\ref{cosineproduct_infinity2} in Eq.\,\ref{eq:r2_productcosine_appendix}, the squared end to end distance normalized can be rewritten in the following form
\begin{equation}
\frac{\left\langle r^{2}\right\rangle }{(N-1)b^{2}}=1+\frac{2}{N-1}\sum_{k=1}^{N-2}\left(N-1-k\right)A_{k}=1+2\sum_{k=1}^{N-2}A_{k}-\frac{2}{N-1}\sum_{k=1}^{N-2}kA_{k}
\end{equation}
where we have reformulated the indexes according to $k=j-i$, as the terms $A_{k}=\left[\mathbf{b}\mathbf{\tilde{T}_{\mathrm{cos}}}^{k}\mathbf{a^{T}}\right]/\lambda_{max}^{k}$ only depend on the difference between $i$ and $j$. Evaluating these $A_{k}$ terms, performing the sums and applying the limit $N\rightarrow\infty$ we get that
\begin{equation}
\frac{\left\langle r^{2}\right\rangle }{(N-1)b^{2}}=1+\frac{2\alpha}{\tilde{u}_{\mathrm{n}}}b_{2}a_{3}+\frac{2\alpha}{1-\alpha}b_{4}a_{4}
\end{equation}
where $\alpha=\left(z\tilde{u}_{\mathrm{n}}\left\langle \cos\gamma\right\rangle\tilde{u}_{\mathrm{nn}}\right)/\lambda_{max}$. Now, replacing the vector elements with the relations from Eq.\,\ref{interaction functions}, one finds that
\begin{equation}
\frac{\left\langle r^{2}\right\rangle }{(N-1)b^{2}}=1+2\left\langle \cos\gamma\right\rangle\left[\varphi_{\bullet\circ\bullet}+\frac{1}{1-\alpha}\xi_{\bullet\bullet\bullet}\right]
\label{e2e_infinity_appendix}
\end{equation}
\end{widetext}

\end{document}